\def\inArXiv{defined}
\ifx\inthesis\undefined
  \ifx\inArXiv\undefined
    \ifx\inIEEE\undefined
      \ifx\inDCC\undefined
        \documentclass[10pt,a4paper]{article} 
        \usepackage[authoryear]{natbib}
        \bibpunct{(}{)}{;}{a}{}{,}
      \else
        \documentclass[smallcaptions,smallabstract]{../style/dccpaper} 
        \usepackage[numbers]{natbib}
        \makeatletter
        \renewcommand{\@biblabel}[1]{[#1]}
        \makeatother
        \makeatletter
        \g@addto@macro\normalsize{%
          \setlength\abovedisplayskip{0.5em}
          \setlength\belowdisplayskip{0.5em}
          \setlength\abovedisplayshortskip{-0.1em}
          \setlength\belowdisplayshortskip{0.5em}
        }
        \g@addto@macro\small{%
          \setlength\abovedisplayskip{0.4em}
          \setlength\belowdisplayskip{0.4em}
          \setlength\abovedisplayshortskip{-0.1em}
          \setlength\belowdisplayshortskip{0.4em}
        }
        \makeatother
      \fi
    \else
      \documentclass[11pt,journal]{../style/IEEEtran}  
      \ifx\useIEEEbib\undefined
        \usepackage[authoryear]{natbib}
      \else
        \usepackage[numbers]{natbib}
      \fi
    \fi
  \else
    \documentclass[11pt,letterpaper,journal]{IEEEtran} 
    \usepackage[authoryear]{natbib}
    \bibpunct{(}{)}{;}{a}{}{,}
  \fi
  \usepackage[utf8]{inputenc}
  \usepackage[russian,english]{babel}
  \usepackage{mthesis}
  \ifx\inDCC\undefined
    \usepackage{fancyhdr}
  \fi
  \DeclareFontShape{OT1}{cmtt}{bx}{n}{<5><6><7><8><9><10><10.95><12><14.4><17.28><20.74><24.88>cmttb10}{}
  \usepackage{etoolbox}
  \usepackage{tikz}
  \usetikzlibrary{shapes,shadows,arrows,calc}
  \usepackage{theorem}
  \usepackage{graphicx}
  \usepackage{subfig}
  \usepackage{makeidx}
  \usepackage{listings}
  \usepackage{msyntax}
  \def\introduce#1{\emph{#1}}
  
\begin{document}
  \def\pieceofwriting{article}
  \def\segment{article}
\else
  \def\segment{chapter}
  \label{msofseq.tex:begin}
\fi

\newif\ifIEEE
\ifx\IEEEkeywords\undefined
  \IEEEfalse
\else
  \IEEEtrue
\fi

\ifx\inDCC\undefined
  \def\thetitle{Compressing Sets and Multisets of Sequences}
\else
  \def\thetitle{\textbf{Compressing Sets and Multisets of Sequences}}
\fi

\def\algname{Algorithm}
\def\Algname{Algorithm}
\def\selfterm{self-delimiting}
\def\Selfterm{Self-delimiting}

\newif\iftwocolumn
\makeatletter
\if@twocolumn
  \twocolumntrue
\else
  \twocolumnfalse
\fi
\makeatother

\ifx\inDCC\undefined
  \def\maybesmall{}
  \def\maybedense{}%
  \def\DCCvspace#1{}
\else
  \def\maybesmall{\renewcommand{\baselinestretch}{0.925}%
                  \small}
  \def\maybedense{\setlength{\itemsep}{-0.125cm}}%
  \def\DCCvspace#1{\vspace*{#1}}
  %
  \makeatletter
  \if@smallcaptions
    \newsavebox{\dcctempbox}
    \renewcommand{\@makecaption}[2]
    {\vspace{10pt}\renewcommand{\baselinestretch}{\smallstretch}
     \small\sbox{\dcctempbox}{#1: #2}
     \ifthenelse{\lengthtest{\wd\dcctempbox > \linewidth}}
     { #1: #2\par}{\begin{center}#1: #2\end{center}}}
  \fi
  \makeatother
\fi

\ifx\inthesis\undefined
  \title{\thetitle}
  \author{Christian Steinruecken\\ University of Cambridge}
  \maketitle
\else
  \chapter{\thetitle}\label{ch:msofseq}%
  \numberwithin{example}{chapter}%
\fi


\def\n[#1]{\ensuremath{n_{\sym{#1}}}}%
\def\TT{\mbox{\tiny\textsf{T}}}%
\def\nt{\ensuremath{n_{\TT}}}%

\makedistvar{DD}{D}%
\makedistvar{LL}{L}%
\makedistvar{PP}{P}%
\makemultiset{M}{m}{M} 
\makemultiset{W}{w}{W} 
\def\LLomit#1#2{\LL[\setminus#1]{#2}}%

%
%
%
%

%
\def\componentwise{com\-po\-nent-wise}%
\def\nobreakhyphen{\raise0.1ex\hbox{-}}%
\def\SHA#1{SHA\nobreakhyphen{}\kern-0.1ex#1}%

%
%
%

\ifx\inthesis\undefined
\fi

\index{multisets|(}%

%


\begin{abstract}
This article describes lossless compression algorithms
for multisets of sequences,
taking advantage of the multiset's unordered structure.
%
Multisets are a generalisation of sets
where members are allowed to occur multiple times.
%
A multiset can be encoded na\"ively by simply storing its elements
in some sequential order, but then information is wasted on
the ordering.
%
We propose a technique
that transforms the multiset
into an order-invariant tree representation,
and derive an arithmetic code that optimally
compresses the tree.
%
Our method
achieves compression
%
even if the sequences in the multiset are individually
incompressible (such as cryptographic
hash sums).
%
%
%
The algorithm is demonstrated practically
by compressing collections of \SHA1\ hash sums,
and multisets of arbitrary, individually encodable
objects.

%
%
\end{abstract}

\ifx\IEEEkeywords\undefined
\else
\begin{IEEEkeywords}
Data compression, source coding, multisets, hash sums, arithmetic coding
\end{IEEEkeywords}
\fi

\section{Introduction}

\ifx\IEEEPARstart\undefined This
\else\IEEEPARstart{T}{his} \fi
article describes a compression algorithm for
multisets of binary sequences that exploits the
disordered nature of the multisets.

%
%
Consider a collection $\setW$ of $N$ words $\bset{\vc{w}{1}{N}}$,
each composed of a finite sequence of symbols.
The members of $\setW$ have no particular ordering
(the labels $w_n$ are used here just to describe the method).
Such collections occur in e.g.~\introduce{bag-of-words} models.
The goal is to compress this collection in such a way
that no information is wasted on the ordering of the words.
%

%
%

Making an order-invariant representation of $\setW$ could be
as easy as arranging the words in some sorted order:
if both the sender and receiver use the same ordering,
zero probability could be given to all words
whose appearance violates the agreed order,
reallocating the excluded probability mass to words
that remain compatible with the ordering.
%
However, the correct probability for
the next element in a sorted sequence is 
expensive to compute,
making this approach unappealing.


%
It may seem surprising at first that a collection of strings can be
compressed
in a way that
does \emph{not} involve encoding or decoding the strings
in any particular order.
The solution presented in this \segment\ is to store them ``all at once''
by transforming the collection to an order-invariant tree representation
and then compressing the tree. 

An example of this technique is presented for collections of sequences
that are independently and identically distributed.
%
The resulting compressing method is demonstrated practically
for two applications:
(1) compressing collections of \SHA1\ sums; and
(2) compressing collections of arbitrary, individually encodable objects.






This is not the first time order-invariant source coding
methods have been considered.
The \textit{bits-back coding} approach
puts wasted bandwidth
to good use by filling it up with \textit{additional data}
\citep{wallace1990a,hinton1994a,frey1997a}.
%
However, it does not solve the problem of compactly encoding only
the desired object.
Much more generally, \citet{varshney2006b,varshney2006a,varshney2007a}
motivate a source coding theory for compressing
sets and multisets.
%
\Citet{reznik2011a} gives a concrete algorithm
for compressing \emph{sets} of sequences,
also with a tree as latent representation,
using an enumerative code \citep{zaks1980a,cover1973a}
for compressing the tree shape.
%
%
Noting that \citeauthor{reznik2011a}'s construction
isn't fully order-invariant,
\citet{gripon2012a} propose
%
%
a slightly more general tree\kern.01em{}-\kern-0.1em{}based
coding scheme for multisets.
Our paper offers a different approach:
we derive the exact distribution over multisets
from the distribution over source sequences,
and factorise it into conditional univariate 
distributions that can be encoded with an arithmetic coder.
%
%
We also give an adaptive, universal code for the case
that the exact distribution 
over sequences is unknown.
%
%
%
%
%


\section{Collections of fixed-length binary sequences}
\index{SHA-1@\SHA{1}|(}

Suppose we want to store a multiset of fixed length binary strings,
for example a collection of hash sums.
The \SHA1\ algorithm \citep{nist1995a} is a file hashing method
which, given any input file,
produces a rapidly computable,
cryptographic hash sum
whose length is exactly 160~bits.
%
Cryptographic hashing algorithms are designed to make it
computationally infeasible to change an input file
without also changing its hash sum.
%
Individual hash sums can be used, for example, to detect
if a previously hashed file has been modified
(with negligible probability of error),
and collections of hash sums can be used
to detect if a given
file is one of a preselected collection of input files.%
\footnote{%
  If an application cares mainly about testing membership
  in a collection,
  even more compact methods exist,
  for example Bloom filters \citep{bloom1970a}.
  Bloom filters are appropriate when a not-so-negligible chance
  of false positives is acceptable.
  %
}

Each bit digit in a random \SHA1\ hash sum is uniformly distributed,
which renders single \SHA1\ sums incompressible.
%
It might therefore seem intuitive at first
that storing $N$ hash sums would cost exactly
$N$ times as much as storing one hash sum.
%
However,
%
%
an \emph{unordered} collection of \SHA1\ sums
can in fact be stored more compactly.
%
For example, the practical savings for a collection of
5000 \SHA1\ sums amount to 10~bits per \SHA1\ sum,
i.e.~each \SHA1\ sum in the collection
takes only 150~bits of space (rather than 160~bits).
The potential saving for a collection of $N$ random hash sums
is roughly $\log_2 N!$ bits.
%

%


A concrete method for compressing multisets of
fixed-length bit strings
(such as collections of \SHA1\ sums)
is described below.
%
%
%
%
%
%
%
%
%
%
%
%
%
%
The algorithm makes use of arithmetic coding
to encode values from {binomial}
and {Beta-binomial} distributions;
%
details
are described in appendices \ref{sec:binomial-code}
and~\ref{sec:betabin-code}.

\subsection{Tree representation for multisets of fixed-length strings}%
\label{sec:fbstree-struc}%
\index{binary trees|(}%

A multiset of binary sequences can be
represented with a binary tree whose nodes store positive
integers.
Each node in the binary tree partitions the multiset
of sequences into two submultisets:
those sequences whose next symbol is a \sym{0},
and those whose next symbol is a \sym{1}.
The integer count $n$ stored in the root node represents
the total size of the multiset,
and the counts $\n[0], \n[1]$ stored
in the child nodes indicate the sizes of their submultisets.
An example of such a tree and its corresponding multiset
is shown in \figname~\ref{fig:fbstree}.

\ifx\inDCC\undefined
  \begin{figure}[t]
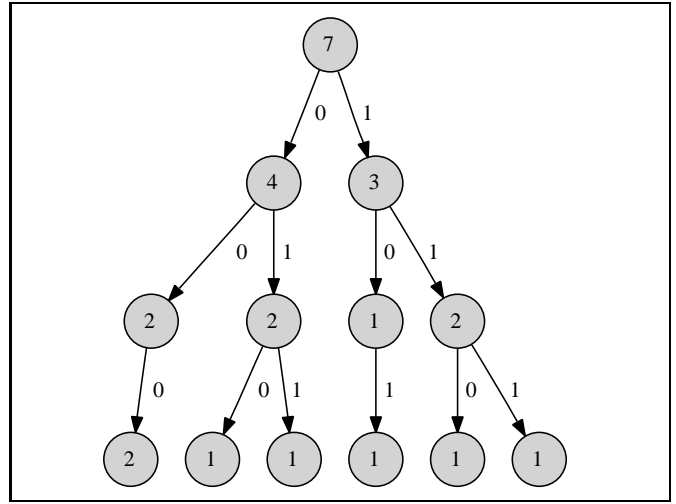

  \iftwocolumn%
    \framebox{%
      \begin{minipage}{0.95\columnwidth}
      \ifx\inArXiv\undefined
        \centerline{\includegraphics[width=0.75\columnwidth]{../graph/bstree1}}
      \else
        \centerline{\includegraphics[width=0.75\columnwidth]{bstree1}}
      \fi
      \end{minipage}
  }
  \else
    \ifx\inArXiv\undefined
      \centerline{\includegraphics[width=0.4\columnwidth]{../graph/bstree1}}
    \else
      \centerline{\includegraphics[width=0.4\columnwidth]{bstree1}}
    \fi
  \fi
  \caption[Binary tree representing a multiset of bit strings.]{
    The binary tree representing the multiset:
    \{~\sym{000}, \sym{000}, \sym{010}, \sym{011},
       \sym{101}, \sym{110}, \sym{111}~\}.
    The count at each node indicates the number of
    strings starting with the node's prefix:
    e.g.~there are
    7~strings starting with the empty string,
    4~strings starting with~\sym{0},
    3~strings starting with~\sym{1}, and
    2~strings starting with~\sym{01}.
  } 
  \label{fig:fbstree}
  \end{figure}
\else
  \begin{figure*}[t]
  \framebox{%
  \begin{minipage}{0.28\textwidth}
    \centerline{\includegraphics[width=0.98\textwidth]{../graph/bstree1}}
  \end{minipage}%
  \begin{minipage}{0.71\textwidth}\centering
  \begin{minipage}{0.92\textwidth}\maybesmall%
    \refstepcounter{figure}\label{fig:fbstree}%
    {Figure~\thefigure: }
    The binary tree representing the multiset:
    \def\sp{,\kern0.5em}
    \begin{center}
      \bset{ \sym{000}\sp \sym{000}\sp \sym{010}\sp \sym{011}\sp
         \sym{101}\sp \sym{110}\sp \sym{111} }.
    \end{center}
    The count at each node indicates the number of
    strings starting with the node's prefix:
    e.g.~there are
    7~strings starting with the empty string,
    4~strings starting with~\sym{0},
    3~strings starting with~\sym{1}, and
    2~strings starting with~\sym{01}.
  \end{minipage}%
  \end{minipage}}
  \vspace*{-0.6em}
  \end{figure*}
\fi

To save space, nodes with zero counts may be omitted from the tree.
For a multiset of fixed-length sequences,
sequence termination is indicated by a leaf node,
or a node that only has children with a count of zero.
%
%
The sequence of branching decisions taken to reach any given
node from the root is called the node's \emph{prefix}.
To recover the original multiset
from the tree,
it suffices to collect the prefix of each leaf node,
including each prefix as many times
as indicated by the leaf node's count.

%
%
%

%

%
%

%
%

A binary tree as described above is unique for any
given collection of binary strings.
The tree can be constructed incrementally,
and supports addition, deletion and membership testing
of sequences in $O(L)$ time, where $L$ is the
sequence length.
%
Merging two trees can be done more efficiently
than adding one tree's sequences to the other individually:
  the counts of nodes whose prefixes are equal can simply
  be added, and branches missing from one tree can be
  copied (or moved) from the other tree.
%
Extracting $N$ sequences from the tree, either
lexicographically or uniformly at random,
takes $O(L \mul N)$ time.

%
%

\subsection{Fixed-depth multiset tree compression algorithm}
\label{sec:fbstree-compr}
\def\nntree{%
  \begin{tikzpicture}[->,baseline=-1.5em,scale=1]\tiny
  \begin{scope}[thin, arr/.style={->,thick,>=latex},
                level/.style={sibling distance=3em, level distance=3.5em},
                comp/.style={circle,draw,fill=gray!30,text centered,
                             text width=1em}]
    \node [comp]  (np)  {$n$}
    child { node [comp] (n0) {$n_0$}
              edge from parent node[yshift=0.4em,xshift=-0.5em] {\sym{0}}
    }
    child { node [comp] (n1) {$n_1$}
              edge from parent node[yshift=0.4em,xshift=+0.5em] {\sym{1}}
    };
  \end{scope}
  \end{tikzpicture}
}%

%
The previous section showed how a multiset of $N$
binary sequences of fixed length $L$ can be converted
to a tree representation.
%
This section
derives exact conditional probability distributions
for the node counts in the resulting tree,
and shows how the tree can
be compactly encoded with an arithmetic coder.
%

Suppose that $N$ and $L$ are known in advance.
With the exception of the leaf nodes,
the count $n$ at any given node in the tree equals
the sum of the counts of its children,
i.e.~$n = \n[0] + \n[1]$.
If the bits of each string are independent and
identically distributed,
the counts of the child nodes (conditional on their parent's count)
jointly follow a binomial distribution:%
\index{binomial distribution}
\begin{equation}\label{eq:bstreenode}
\begin{array}{ccc}
 \begin{array}{rcl}
    n_\sym{1} &\sim& \Binomial{\textstyle n,\ \theta} \\
    n_\sym{0} &=& n - n_\sym{1}
  \end{array}
& \hspace{3em}
& \nntree
\end{array}
\end{equation}
where $\theta$ is the probability of symbol $\sym{1}$.
If the symbols $\sym{0}$ and $\sym{1}$ are uniformly
distributed (as is the case for \SHA1\ sums),
$\theta$ should be set to $\frac12$.
Given the parent count $n$, only one of the child
counts needs to be communicated, as the other
can be determined by subtraction from $n$.
%
%
%
Since all strings in the multiset have length $L$,
all the leaf nodes in the tree are located at depth $L$,
making it unnecessary to communicate which of
the nodes are leaves.

If $N$ and $L$ are known, the tree can be communicated
as follows:
Traverse the tree, except for the leaf nodes,
starting from the root (whose count $N$ is already known).
%
Encode one of child counts (e.g.~$\n[1]$) using
\ifx\inthesis\undefined
a binomial code
\else
a binomial code (described in section~\ref{sec:binomial-code}),
\fi
\index{binomial code}
and recurse on all child nodes whose count
is greater than zero.
The parameters of the binomial code are
the count of the parent, and the symbol bias $\theta$,
as shown in equation~\eqref{eq:bstreenode}.
%
%
The tree can be traversed in any order that
visits parents before their children.

This encoding process is invertible, allowing perfect recovery
of the tree.  The same traversal order must be followed,
and both $N$ and $L$ must be known
(to recover the root node's count, and to determine which
nodes are leaf nodes).
Depending on the application, $N$ or $L$ can be transmitted
first using an appropriate code
\ifx\inthesis\undefined
over integers.
\else
over integers, such as in section~\ref{sec:integer-codes}.
\fi
%
A concrete coding procedure
using pre-order traversal is shown in
\algname~\ref{alg:fbstree}. 

\def\rootof#1{\ensuremath{\text{\texttt{root}}\left(#1\right)}}
\def\isleaf#1{\ensuremath{\text{\texttt{is\_leaf}}\left(#1\right)}}
\def\us{\kern-0.1ex{}\_\kern0.2ex{}}
\def\encodeNode#1{\ensuremath{\text{\texttt{encode\us{}node}}\!\left(#1\right)}}
\def\decodeNode#1{\ensuremath{\text{\texttt{decode\us{}node}}\!\left(#1\right)}}
%
\ifx\inDCC\undefined\else
  \setalgBLUE
  \def\algheaderstyle#1{\textbf{#1}}
\fi%
\begin{algorithm*}[tpb]
\maybesmall%
\medskip
{\algheader{Coding algorithm for multisets of fixed-length sequences}}
\begin{minipage}{0.49\textwidth}
\centerline{\algsubheader{Encoding}}
\end{minipage}%
\begin{minipage}{0.49\textwidth}
\centerline{\algsubheader{Decoding}}
\end{minipage}%
\medskip

\begin{minipage}[t]{0.49\textwidth}
\textbf{Inputs:} $L$,\ binary tree $T$
\begin{enumerate}[A.]
\item Encode $N$, the count of $T$'s root node,
      using a code over positive integers.
\item Call \encodeNode{T}.
\end{enumerate}
\end{minipage}%
\hspace*{0.02\textwidth}%
\begin{minipage}[t]{0.49\textwidth}
\textbf{Input:} $L$ \quad \textbf{Output:} binary tree $T$
\begin{enumerate}[A.]
\item Decode $N$, using the same code over positive integers.
\item Return $T \assign \decodeNode{N,L}$.
\end{enumerate}%
\end{minipage}
\medskip
\algrule
\medskip
\begin{minipage}[t]{0.49\textwidth}
\textbf{subroutine} \encodeNode{t}:\\
\hspace*{0.07\columnwidth}%
\begin{minipage}[t]{0.93\columnwidth}
  \vspace{-\smallskipamount}
  If node $t$ is a leaf:
  \begin{enumerate}\maybedense
  \item Return.
  \end{enumerate}
  Otherwise:
  \begin{enumerate}\maybedense
  \item Let $t_{\sym{0}}$ and $t_{\sym{1}}$ denote the children of $t$,
        and $\n[0]$ and $\n[1]$ the children's counts.
  \item Encode $\n[1]$ using a binomial code,
        as $\n[1] \sim \Binomial{\n[0]+\n[1],\ \theta}$.
  \item If $\n[0] > 0$, call $\encodeNode{t_{\sym{0}}}$.
  \item If $\n[1] > 0$, call $\encodeNode{t_{\sym{1}}}$.
  \end{enumerate}
\end{minipage}
\end{minipage}
\hspace*{0.02\textwidth}%
\begin{minipage}[t]{0.49\textwidth}
\textbf{subroutine} \decodeNode{n,l}:\\
\hspace*{0.07\columnwidth}%
\begin{minipage}[t]{0.93\columnwidth}
  \vspace{-\smallskipamount}
  If $l > 0$ then:
  \begin{enumerate}\maybedense
  \item Decode $\n[1]$ using a binomial code,\\
        as $\n[1] \sim \Binomial{n,\ \theta}$.
  \item Recover $\n[0] \assign \left( n - \n[1] \right)$.
  \item If $\n[0] > 0$, then:\\
        \hspace*{0.8em} $t_{\sym{0}} \assign \decodeNode{\n[0], l-1}$.
  \item If $\n[1] > 0$, then:\\
        \hspace*{0.8em} $t_{\sym{1}} \assign \decodeNode{\n[1], l-1}$.
  \item Return a new tree node with count $n$ and
        children $t_{\sym{0}}$ and $t_{\sym{1}}$.
  \end{enumerate}
  Otherwise, return null.
\end{minipage}
\end{minipage}
\algfooter
\caption[Coding algorithm for multisets of fixed-length sequences]{%
  \maybesmall%
  Coding algorithm for binary trees representing multisets of
  binary sequences of length $L$.
  The form and construction of the binary tree
  are described in section~\ref{sec:fbstree-struc}.
  Each tree node $t$ contains an integer count $n$
  and two child pointers $t_{\sym{0}}$ and $t_{\sym{1}}$.
  %
  The counts of the children are written
  $\n[0]$ and $\n[1]$.
  If $\n[0]$ and $\n[1]$ are zero, $t$ is deemed to be a leaf,
  and vice versa.
  %
  %
  %
  %
  %
  %
  %
  %
  $T$ denotes the tree's root node.
  %
  %
  \index{binomial code}
}\label{alg:fbstree}
\end{algorithm*}

%
%
%

%
%



\medskip

\textbf{Application to \SHA1\ sums. }
For a collection of $N$ \SHA1\ sums, the depth of the binary
tree is $L=160$, and the root node contains the integer $N$.
%
%
%
%
%
If the \SHA1\ sums in the collection are random and
do not repeat,
the distribution over the individual bits in each
sequence is uniform,
making a binomial code with bias $\theta=\frac12$
an optimal choice.
However, if the collection is expected to contain
duplicates,
the distribution over the counts is no longer
binomial with a fixed bias;
in fact, the bias might then be different for each
node in the tree.
%
%
In such a case, a \Betabinomial\ code may be more
appropriate, as it can \emph{learn} the underlying symbol
probability $\theta$ independently for each node,
rather than assuming it to have a particular fixed value:
\begin{equation}\label{eq:betadist}
 \begin{array}{ccl}
    n_\sym{1} &\sim& \BetaBin{\textstyle n, \alpha, \beta} \\
    n_\sym{0} &=& n - n_\sym{1}
  \end{array}
\end{equation}
%
%
\ifx\inDCC\undefined
A \Betabinomial\ coding procedure is described
in appendix~\ref{sec:betabin-code}.

\fi
%
%
%
%
%
The tree coding method of \algname~\ref{alg:fbstree}
can be modified to use a \Betabinomial\ code
by replacing the encoding and decoding calls in
the subroutine accordingly.
%
In our experiments, 
the \Betabinomial\ parameters were set to $\alpha=\frac12$
and $\beta=\frac12$.

\begin{figure*}
\ifx\inArXiv\undefined
  \smartgraphics{23}{410}{0}{9}{149}{23}%
     {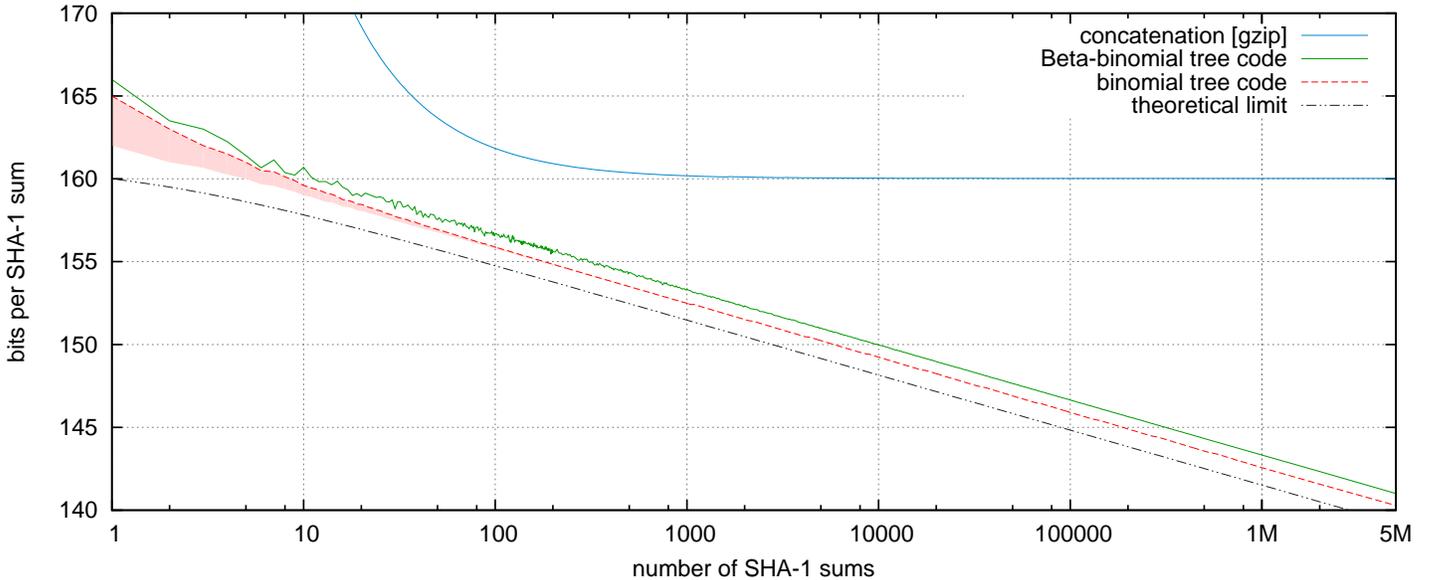}%
     {}{}%
\else
  \smartgraphics{23}{410}{0}{9}{149}{23}%
     {bpn_rsha1_wide.eps}%
     {}{}%
\fi
\DCCvspace{-0.9\bigskipamount}
\caption[Compressing multisets of \SHA1\ sums]{%
  Practical lossless compression performance
  of the fixed-depth multiset tree compressor
  on multisets of \SHA1\ sums.
  %
  %
  %
  %
  For each position on the x-axis,
  $N$ uniformly distributed 64-bit random numbers were generated
  and hashed with \SHA1; the resulting multiset of $N$ \SHA1\ sums
  was then compressed with each algorithm.
  The winning compression method is \algname~\ref{alg:fbstree}
  using a binomial code,
  where $N$ itself is encoded with a Fibonacci code.\index{Fibonacci code}
  The shaded region
  indicates the proportion of information used by the Fibonacci
  code. 
  The theoretical limit is $160 - \frac1{N} \log_2 N!$ bits.
  For comparison, \progname{gzip} was used to compress
  the concatenation of the $N$ \SHA1\ sums;
  reaching, as expected, 160 bits per \SHA1\ sum.
  %
}\label{fig:rsha1}
\end{figure*}

The practical performance of the algorithm on
multisets of \SHA1\ sums is
shown in \figname~\ref{fig:rsha1}.
%
The multisets used in this experiment contain no
duplicate hashes, so the compression achieved by
the algorithm really results from exploiting the
permutation invariance of the multiset
rather than redundancy in the hashes.
%

%
%

\index{SHA-1@\SHA{1}|)}

\section{Collections of binary sequences of arbitrary length}

\ifx\inDCC\undefined
  The method of the previous section transformed a collection
  of fixed-length binary sequences into a binary tree, and described
  a compression method for storing the tree in a space-efficient way.
  The property that the sequences in the collection had
  the same length $L$ was
  a prerequisite for the method to work.
  In this section, the method is generalised to admit
  binary sequences of arbitrary length.
\else
  This section generalises the tree coding method 
  to admit binary sequences of arbitrary length.
\fi
Two approaches are considered for encoding the termination
of sequences in the tree:
the first approach covers collections of \selfterm\
sequences, which allow the tree to be compressed without
encoding additional information about termination.
The second approach, for arbitrary sequences,
assumes a \emph{distribution} over sequence lengths
and encodes sequence termination directly in the tree nodes.
%
For either approach,
the same binary tree structure is used as before,
except that
sequences stored in the tree can now have any length.
%

\subsection{Compressing multisets of \selfterm\ sequences}

%
%
%
\Selfterm\ sequences encode their own length,
i.e.~it can be determined from the sequence itself
if further symbols follow or if the sequence has ended.
Many existing compression algorithms produce \selfterm\ sequences,
e.g.~the Huffman algorithm, codes for integers,
or suitably defined arithmetic coding schemes.
%
%
%
A multiset of such \selfterm\ sequences has the property
that,
for any two distinct sequences in the multiset,
neither can be a prefix of the other.

%

Consider the tree corresponding
to such a multiset of binary strings.
Because of the prefix property,
all sequences in the tree will terminate at leaf nodes,
and the counters stored in child nodes always
add up to the counter of the parent node.
%
Consequently, the same compression technique can be used
as for fixed-length sequences.
\Algname~\ref{alg:fbstree} applies exactly as before,
with the exception that the end-of-string detector in
the decoder must be modified to detect
the end of each \selfterm\ sequence.

\begin{figure*}
\ifx\inArXiv\undefined
  \smartgraphics{25}{408}{0}{9}{149}{23}
     {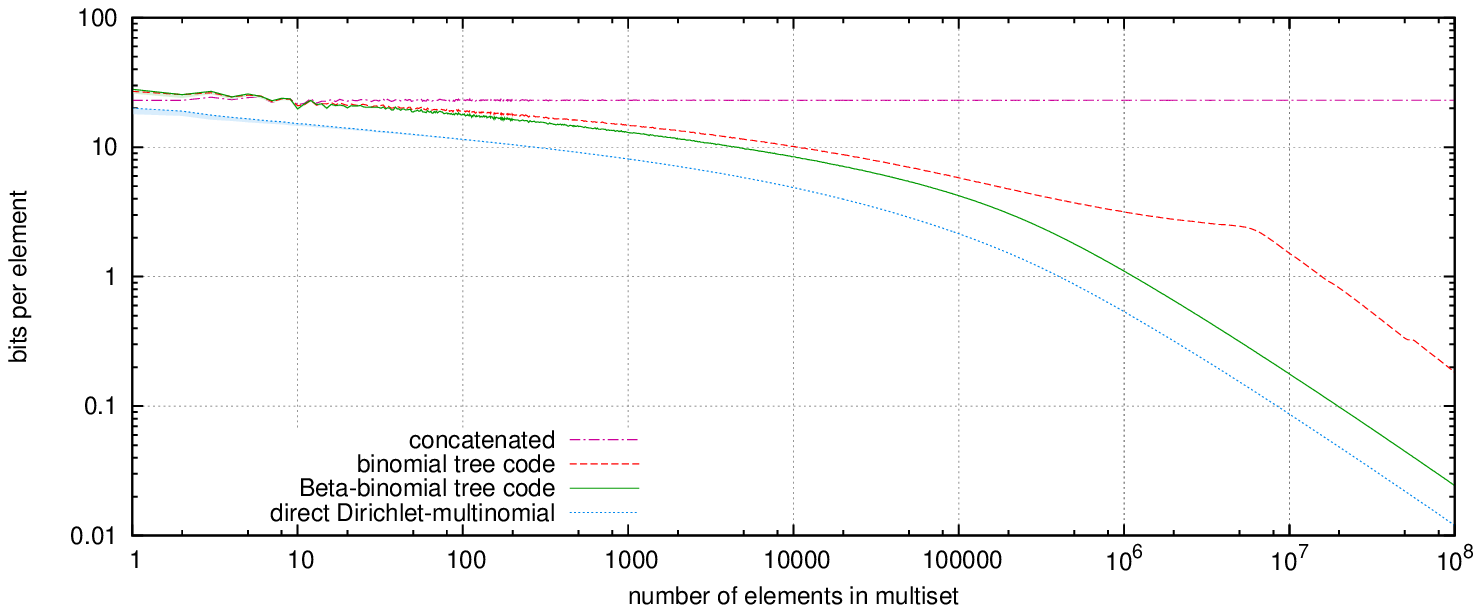}%
     {}{}%
\else
  \smartgraphics{25}{408}{0}{9}{149}{23}
     {bpn_rnd100k_fibs_wide.eps}%
     {}{}%
\fi
%
%
\DCCvspace{-0.9\bigskipamount}
\caption[Compressing multisets of self-terminating sequences]{%
  Experimental compression performance of various algorithms
  on multisets of \selfterm\ sequences.
  For each position on the x-axis, a multiset of $N$ \selfterm\ 
  sequences was generated by taking $N$ uniformly distributed
  integers between 1 and 100\ns000 and encoding each number with a
  Fibonacci code.\index{Fibonacci code}
  The multiset of the resulting code words
  was then compressed with each algorithm.
  The y-axis shows the compressed size in bits divided by $N$.
  %
  %
  %
  The flat concatenation of the sequences in the multiset
  is included for reference (achieving zero compression).
  For comparison, the source multisets of integers
  (rather than the multisets of Fibonacci-encoded integers)
  were compressed directly with a Dirichlet-multinomial code.
  The (barely visible) shaded regions indicate the amount of
  information taken up by the Fibonacci code to encode $N$ itself.
  %
  \graphindex{Fibonacci code}
}\label{fig:ms-rnd100k-fibs}
\end{figure*}

\textbf{Compressing arbitrary multisets. }
Consider a random multiset $\setM$
over an arbitrary space $\setX$,
whose elements can be independently
compressed to \selfterm\ binary strings (and reconstructed
from them).
%
%
%
Any such multiset $\setM$
can be losslessly and reversibly converted
to a multiset $\setW$ of \selfterm\ sequences,
and $\setW$ can be compressed and decompressed
with the tree coding method as described above.
%
%
%
%
%
%

\textbf{Alternative.}
A random multiset $\setM$ is most effectively compressed with a compression
algorithm that exactly matches $\setM$'s probability distribution;
we'll call such an algorithm a \emph{direct code} for $\setM$.
%
%
When a direct code is not available or convenient,
the indirect method of first mapping $\setM$ to $\setW$
might be a suitable alternative.

\textbf{Experiment.}
Experimental results of this approach on random
multisets of \selfterm\ sequences are shown
in \figname~\ref{fig:ms-rnd100k-fibs}.
%
Each multiset was generated by drawing $N$ uniform
random integers and converting these integers to
\selfterm\ sequences with
a Fibonacci code\index{Fibonacci code}
\citep{kautz1965a,apostolico1987a}.%
\footnote{%
    The Fibonacci code was chosen for elegance.
    However, any code over integers could be used,
    e.g.~an exponential Golomb code \citep{teuhola1978a}
    or the $\omega$-code by \citet{elias1975a}.
}
The Beta-binomial variant of the tree coder
wins over the binomial variant, and closely follows
the trajectory of a Dirichlet-multinomial code
for the underlying multisets of integers.
\ifx\inDCC\undefined%
  A brief description of the Fibonacci code
  can be found in appendix~\ref{sec:fibcode}.
\fi


\subsection{Encoding string termination via end-of-sequence markers}
%

\begin{figure*}[t]
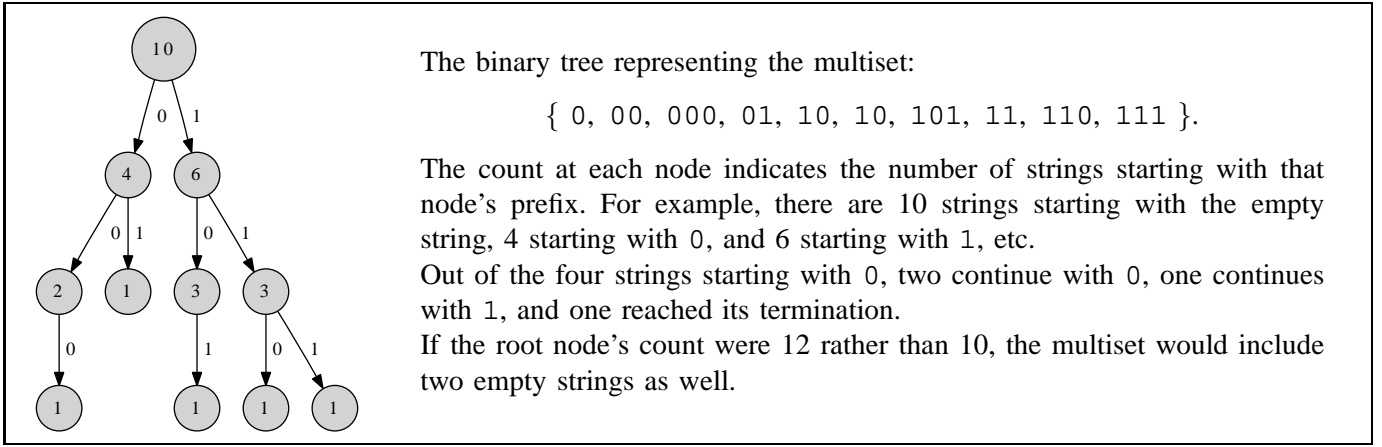

\framebox{%
\begin{minipage}{0.27\textwidth}
\begin{center}
\ifx\inArXiv\undefined
  \includegraphics[width=0.9\textwidth]{../graph/vbtree1}
\else
  \includegraphics[width=0.9\textwidth]{vbtree1}
\fi
\end{center}
\end{minipage}%
\begin{minipage}{0.72\textwidth}\centering
\begin{minipage}{0.92\textwidth}\maybesmall%
The binary tree representing the multiset:
\def\sp{,\kern0.5em}
\begin{center}
\bset{ \sym{0}\sp \sym{00}\sp \sym{000}\sp
\sym{01}\sp
\sym{10}\sp \sym{10}\sp \sym{101}\sp
\sym{11}\sp
\sym{110}\sp
\sym{111} }.
\end{center}
The count at each node indicates the number of
strings starting with that node's prefix.
For example, there are
10~strings starting with the empty string,
4~starting with~\sym{0},
and 6~starting with~\sym{1}, etc.

Out of the four strings starting with~\sym{0},
two continue with~\sym{0}, one continues with~\sym{1},
and one reached its termination.

If the root node's count were 12 rather than 10,
the multiset would include two empty strings as well.

\end{minipage}
\end{minipage}}
\caption[Binary tree representing a multiset of binary strings.]{%
  Binary tree representing a multiset of ten
  binary sequences. 
  %
  This tree follows the same basic structure as the tree in
  \figname~\ref{fig:fbstree}, but admits sequences of
  variable length.
  The tree representation is unique for each multiset.
}\label{fig:vbtree}
\end{figure*}

Consider now a multiset containing binary sequences of
arbitrary length,
whose sequences lack the property that
their termination can be determined from a prefix.
This is the most general case.
%
In this scenario, it is possible for the multiset to
contain strings where one is a prefix of the other,
for example \sym{01} and \sym{011}.
To encode such a multiset, string termination must
be communicated explicitly for each string.
Luckily, the existing tree structure can be used as before
to store such multisets; the only difference is that
the count of a node need not equal the sum of the counts
of its children, as terminations may now occur at any node,
not just at leaf nodes.
Both child counts therefore need to be communicated.
An example of such a tree is shown in \figname~\ref{fig:vbtree}.
%

%
The counter $n$ stored in each node still indicates the number
of sequences in the collection that start with that node's prefix.
The number of terminations $\nt$ at any given node
equals the difference of the node's total count $n$
and the sum of its child counts $\n[0]$ and $\n[1]$.

%
%

%
Suppose that the number $N = \setsize{\setW}$
of sequences in the multiset $\setW$ is distributed according
to some distribution $\txtDD$ over positive integers,
and that the length of each sequence $w_n \in \setW$
is distributed according to some distribution $\txtLL$.
Given $\txtDD$ and $\txtLL$,
a near-optimal compression algorithm for the multiset $\setW$
can be constructed as follows.


First, form the tree representation of $\setW$, following
the construction described in the previous section.
%
%
The count of the root node can be communicated
using a code for $\txtDD$.
%
%
Each node in the tree has a count $n$,
child counts $\n[0]$ and $\n[1]$,
and an implicit termination count $\nt$ 
fulfilling $n = \n[0] + \n[1] + \nt$.
Assuming that the bits at the same position of
each sequence are independent and
identically distributed, the values of
$\n[0]$, $\n[1]$ and $\nt$
are multinomially distributed (given $n$).

The parameters of this multinomial distribution
can be derived from $\txtLL$ as follows:
%
%
The $n$ sequences described by the current
node have a minimum length of $d$, where $d$ is the
node's depth in the tree (the root node is located at depth 0).
%
Out of these $n$ sequences,
$\n[0]$ continue with symbol $\sym{0}$,
$\n[1]$ continue with symbol $\sym{1}$, and
$\nt$ terminate here.
%
%
Given the sequence length distribution $\txtLL$,
the probability for a sequence that has at least $d$ symbols
to have no more than $d$ symbols is given by
a Bernoulli distribution with bias $\theta_{\TT}(d)$,
%
where:
\DCCvspace{-\medskipamount}
\begin{equation}
\theta_{\TT}(d)
\ :=\ 
    \displaystyle \frac{\LL{d}}
                       {{1 - \sum_{k<d} \LL{k}}}
\end{equation}
%
%
%
%
%
%
%
%
%
%

%
%
Consequently, the number of terminations $\nt$
at depth $d$
(out of $n$ possible sequences)
is binomially distributed with:%
\index{binomial distribution}
\begin{equation}
     \nt \ \sim\  \Binomial{n,\ \theta_{\TT}(d) } \\
\end{equation}
%
%
Writing $\theta_{\TT}$ for
the probability of termination at the local node,
and $\theta_{\sym{1}}$
and $\theta_{\sym{0}}$
for the occurrence probabilities of $\sym{1}$ and $\sym{0}$,
the joint distribution over $\n[0]$, $\n[1]$ and $\nt$ can
be written as follows:
\begin{equation}
\left( \nt, \n[0], \n[1] \right)
\ \sim\ 
       \Mult{\theta_{\TT},\ 
             \theta_{\sym{0}} \left(1\?-\theta_{\TT}\right),\ 
             \theta_{\sym{1}} \left(1\?-\theta_{\TT}\right)}
\end{equation}
where
$\theta_{\sym{1}} = 1 - \theta_{\sym{0}}$.
%
%
The encoding procedure for this tree needs to
encode a ternary (rather than binary) choice,
but the basic principle of operation remains the same.
\Algname~\ref{alg:fbstree} can be modified to encode
$\left( \nt, \n[0], \n[1] \right)$ using
\ifx\inthesis\undefined
a multinomial code.
\else
a multinomial code,
e.g.~the method described in section~\ref{sec:mult-code}.
\fi
%

Note that, as described above, $\theta_{\TT}$ is
a function of the length distribution $\txtLL$
and the current node depth $d$.
In principle, it is possible to use a conditional length distribution
that depends on the prefix of the node, as the node's
prefix is available to both the encoder and the decoder.
Similarly, $\theta_{\sym{0}}$ and $\theta_{\sym{1}}$
could in principle be functions of depth or prefix.
%
%
%
%
%

\section{Conclusions}

We proposed a novel and simple data compression algorithm
for sets and multisets of sequences, and illustrated its
use on collections of cryptographic hash sums.
Our approach is based on the general principle that
one should encode a permutation-invariant
representation of the data, in this case a tree,
with a code that matches the distribution induced by
the data's generative process.
When the distribution of the source sequences is known,
the tree is optimally compressed with
a nested binomial coding scheme;
otherwise, a Beta-binomial coding scheme
can be used.  The Beta-binomial code is universal
in that it learns the symbol distribution of the sequences
in the multiset
(even for symbol distributions that are position
or prefix dependent).

One might regard the coding algorithms
presented in this paper
either as lossless compression for sets and multisets,
or as lossy compression methods for lists:
when the order of a list of elements isn't important,
bandwidth can be saved.
%

%
Future work could address multisets of sequences whose
elements are not independent and identically distributed,
by combining the above approach with probabilistic models
of the elements.

\index{binary trees|)}


\index{multisets|)}%

\ifIEEE
  \appendices
\else
  \ifx\inDCC\undefined
    \appendix
  \else
    \section*{Appendix}
    \vspace*{-0.8em}
    \appendix
    \begingroup
    \maybesmall
  \fi
\fi

\section{A binomial code}\label{sec:binomial-code}

The binomial distribution describes the number of
successes in a set of $N$ independent Bernoulli
trials.\index{binomial distribution}
It is parametrised by natural number $N$ and
success probability $\theta$,
and ranges over positive integers $n \in \bset{0 \dots N}$.
A~binomial random variable has the
following probability mass function:
\DCCvspace{-\medskipamount}
\begin{equation}
\Binomial{n \| N, \theta} = {N \choose n} \mul \theta^n (1-\theta)^{N-n}
\end{equation}
Encoding a binomial random variable with an arithmetic coder
requires computing the cumulative distribution function
of the binomial distribution.
A method for doing this efficiently might utilise the
following recurrence relation:
\iftwocolumn
  \begin{equation}
  \begin{array}{cl}
    \multicolumn{2}{l}{\Binomial{n+1 \| N, \theta}} \\[0.5em]
    =& \displaystyle
        \frac{N-n}{n+1} \mul \frac{\theta}{1-\theta}
        \mul \Binomial{n \| N, \theta}
  \end{array}
  \end{equation}
\else
  \begin{equation}
    \Binomial{n+1 \| N, \theta}
    \ =\  \frac{N-n}{n+1} \mul \frac{\theta}{1-\theta}
          \mul \Binomial{n \| N, \theta}
  \end{equation}
\fi
The cumulative binomial distribution can then be computed
as follows.
Initialise
$B_{\Sigma} \assign 0$, and $B \assign (1-\theta)^N$.
To encode a binomially distributed value $n$,
repeat for each $k$ from $1 \dots n$:
\DCCvspace{-\bigskipamount}
\begin{eqnarray}
B_{\Sigma} &:=& B_{\Sigma} + B \\
B &:=& \frac{N-k}{k+1} \mul \frac{\theta}{1-\theta} \mul B
\end{eqnarray}
The interval $\left[ B_{\Sigma},\ B_{\Sigma}\?+B \right)$
is then a representation of $n$
that can be used with an arithmetic coder.
%

\section{A Beta-binomial code}\label{sec:betabin-code}
The \Betabinomial\ compound distribution results from
integrating out the success parameter $\theta$
of a binomial distribution, assuming $\theta$ is Beta distributed.
\index{Beta-binomial distribution@\Betabinomial\ distribution}
%
It is
parametrised by an integer $N$ and the parameters
$\alpha$ and $\beta$ of the Beta prior:
\iftwocolumn
  \begin{align}
     & \hspace*{-0.9em} \BetaBin{n \| N, \alpha, \beta}
        \nonumber\\[0.2em]
    =& \displaystyle
       \int \Binomial{n \| N, \theta} \mul
            \Beta{\theta \| \alpha,\beta} \d{\theta} \\
    =& \displaystyle
       {N \choose n}
         \mul \frac{\G{\alpha\?+\beta}}{\G{\alpha} \G{\beta}}
         \mul \frac{\G{\alpha\?+n} \G{\beta\?+N\?-n}}{\G{\alpha\?+\beta\?+N}}
  \end{align}
\else
  \begin{align}
     \BetaBin{n \| N, \alpha, \beta}
    \ =\ & \displaystyle
           \int \Binomial{n \| N, \theta} \mul
                \Beta{\theta \| \alpha,\beta} \d{\theta} \\
    \ =\ & \displaystyle
       {N \choose n}
         \mul \frac{\G{\alpha\?+\beta}}{\G{\alpha} \G{\beta}}
         \mul \frac{\G{\alpha\?+n} \G{\beta\?+N\?-n}}{\G{\alpha\?+\beta\?+N}}
  \end{align}
\fi
%
%
%
Just like for the binomial distribution, there is
a recurrence relation which can speed up the computation
of the cumulative \Betabinomial\ distribution:
\iftwocolumn
  \begin{gather}\label{eq:betabin-rec}
  \begin{array}{cl}
    \multicolumn{2}{l}{\BetaBin{n+1 \| N, \alpha, \beta}} \\[0.5em]
  =& \displaystyle
     \frac{N\?-n}{n\?+1} \mul \frac{\alpha\?+n}{\beta\?+N\?-n\?-1}
     \mul \BetaBin{n   \| N, \alpha, \beta}
  \end{array}\raisetag{4em}
  \end{gather}
\else
  \begin{equation}\label{eq:betabin-rec}
    \BetaBin{n+1 \| N, \alpha, \beta}
    \ =\ 
    \frac{N\?-n}{n\?+1} \mul \frac{\alpha\?+n}{\beta\?+N\?-n\?-1}
    \mul \BetaBin{n   \| N, \alpha, \beta}
  \end{equation}
\fi
The method from appendix~\ref{sec:binomial-code}
can be modified accordingly, yielding a
\Betabinomial\ coding scheme.

%

\ifx\inDCC\undefined

  \section{Fibonacci code for integers}\label{sec:fibcode}
  \begin{table}[h]
  \begin{center}
  \def\fsym#1{\texttt{#1}}%
  \begin{tabular}{rlcrlcrl}
  1  &  \fsym{11}      && 8  &  \fsym{000011}   && 15 &  \fsym{0100011} \\
  2  &  \fsym{011}     && 9  &  \fsym{100011}   && 16 &  \fsym{0010011} \\
  3  &  \fsym{0011}    && 10 &  \fsym{010011}   && 17 &  \fsym{1010011} \\
  4  &  \fsym{1011}    && 11 &  \fsym{001011}   && 18 &  \fsym{0001011} \\
  5  &  \fsym{00011}   && 12 &  \fsym{101011}   && 19 &  \fsym{1001011} \\
  6  &  \fsym{10011}   && 13 &  \fsym{0000011}  && 20 &  \fsym{0101011} \\
  7  &  \fsym{01011}   && 14 &  \fsym{1000011}  && 21 &  \fsym{00000011} \\
  \end{tabular}
  \end{center}
  \caption[Fibonacci code words]{%
    Code words assigned to integers $n=1 \dots 21$
    by a Fibonacci code.
  }\label{tab:fibcode}
  \end{table}
  A Fibonacci code represents any natural number $n$ as a sum of
  Fibonacci numbers,\index{Fibonacci numbers}
  where each Fibonacci number may occur at most once.%
  \footnote{See \citet[Th\'eor\`emes Ia\,\&\,Ib]{zeckendorf1972a}
  for a proof that this representation is unique
  and exists for all integers.}
  \index{Zeckendorf representation}%
  The encoding procedure for $n$ works as follows:
  For each Fibonacci number from $F(2)$ upwards, a \sym{1} is written
  for presence or a \sym{0} for absence of $F(k)$
  in $n$'s Fibonacci sum representation.
  The procedure stops after the \sym{1}-digit for the largest
  contained Fibonacci number is written.
  The resulting code words are unique, and have the additional
  property that the substring \sym{11} never occurs.
  Appending an additional \sym{1} to the encoded number thus
  marks termination.

\fi

\ifIEEE
\else
  \ifx\inDCC\undefined
  \else
     \endgroup
  \fi
\fi

%
%
\ifx\inthesis\undefined
  \ifx\inArXiv\undefined
    \ifx\useIEEEbib\undefined
      \ifx\inDCC\undefined
        \bibliographystyle{mythesis} 
      \else
        \bigskip
        {\maybesmall\noindent
        \textbf{Acknowledgments.}
        The author would like to thank David MacKay,
        Zoubin Ghahramani, Jossy Sayir 
        and anonymous reviewers for helpful feedback.}
        \bibliographystyle{../style/IEEEtranN} 
        \section*{References}
      \fi
    \else
      \bibliographystyle{../style/IEEEtranN}  
    \fi
    \bibliography{../style/IEEEabrv,msofseq}
  \else
    \bigskip
    {\maybesmall\noindent
    \textbf{Acknowledgments.}
    The author would like to thank David MacKay,
    Zoubin Ghahramani and Jossy Sayir 
    and anonymous reviewers for helpful feedback.}
    \ifx\useIEEEbib\undefined
      \bibliographystyle{mythesis}   
    \else
      \bibliographystyle{IEEEtranN}  
    \fi
    \bibliography{IEEEabrv,msofseq}
  \fi
  \end{document}
\else
  \label{msofseq.tex:end}
  \regfilepp{msofseq.tex}{$ $Revision: 1.27 $ $}{$ $Date: 2013/11/12 01:25:12 $ $}{\pageref{msofseq.tex:begin}}{\pageref{msofseq.tex:end}}
\fi